# Prediction of Atomization Energies of $Au_{13}^+$ Clusters through the Machine Learning Approach

*AUTHOR NAMES*

Yasuharu Okamoto

 AUTHOR ADDRESS

Data Platform Center, National Institute for Materials Science, 1-2-1 Sengen, Tsukuba, Ibaraki

305-0047, Japan

**Corresponding Author**

Yasuharu Okamoto

KEYWORDS     machine learning, kernel ridge regression, density functional theory, cluster

ABSTRACT

We examine a new method for predicting the atomization energies of $Au_{13}^+$ clusters by a nonlinear regression model using interatomic and centroid distances as descriptors to improve the efficiency of density-functional theory calculations. Learning data were created using the time-series data of atomic coordinates and Kohn-Sham energy generated by molecular-dynamics simulations. This approach predicted the atomization energies of fifteen known stable/metastable structures of $Au_{13}^+$ clusters well. Moreover, we found that the fitting to the test data could be markedly improved by eliminating the descriptors representing the short interatomic distance.

Introduction

Density-functional theory (DFT) calculations require large computational resources, although they can predict the various properties of materials to high accuracies. To deal with the problem, several strategies have been explored so far: the adoption of algorithms suitable for parallel computation such as the finite difference method with real space grids[1] or the divide-and-conquer method,[2] fragment molecular orbital method,[3] development of a new computing formulation such as the Order-$N$ method,[4,5] or a semi-empirical method based on the parameters determined from ab-initio calculations.[6] With rapid progress in machine learning in recent years, studies to overcome this problem using machine learning approaches have been conducted.[7–15]

There seem to be two considerations in machine learning: the design of descriptors and the reduction of learning cost. Concerning molecular systems, standard molecular fingerprints such as MACCS and FP4 exist in cheminformatics and have been widely used for molecular similarity evaluations in the pharmaceutical field.[16–19] Molecular fingerprints are generally suitable for organic compounds. On the other hand, there are no standard descriptors like molecular fingerprints for crystal/condensed matter systems, and thus, inventing good descriptors has been an important research theme in materials informatics.[20–22]

Because the main input/output data for DFT calculations are the atomic coordinates, it is probably simplest to construct descriptors based on atomic coordinates. It is easy to extend descriptors by including not only structural information based on atomic coordinates but also molecular properties such as the melting and boiling points. However, the data for such properties may not exist for the targeted molecules. Unknown virtual molecules are also targeted in computational molecular design.[23] Thus, it is necessary to predict the properties of such molecules. This makes the feature design complicated. To reduce learning costs, it is important to automatically create various structures suitable for supervised learning without human intervention.

In this study, we examined a new approach combining DFT calculations with machine learning, where the learning data were efficiently generated using time-series data from a molecular-dynamics (MD) simulations, and descriptors were constructed from the information of atomic coordinates only. The new method was applied to the prediction of the atomization energies of $Au_{13}^+$ clusters. The gold clusters, composed of metallic bonds, had flexibility in their bond angles, unlike the clusters made of covalent or ionic bonds. This allowed the gold clusters to have various structures. In addition, $Au_{13}^+$ clusters seemed to be suitable for verifying this approach because Gilb et al. reported 15 stable/metastable structures.[24]

Computational Method

**DFT calculations:** All DFT calculations were done using the Quantum Espresso (ver. 6.1) program package.[25] The generalized gradient approximation (GGA) to the exchange-correlation functional formulated by Perdew, Burke, and Ernzerhof (PBE96) was used in the calculation.[26] Ultrasoft pseudopotential was employed for the Au atom in the calculation.[27] Plane-wave basis sets with cutoff energies of 30 and 300 Ry were used for the expansion of wave functions and charge density, respectively. Gamma-point sampling was used for Brillouin zone (BZ) integration. The $Au_{13}^+$ cluster was placed in a cubic supercell with 30 Bohr side and a uniform background negative charge was added to preserve the charge neutrality of the supercell. MD simulations were carried out at 1500 K using the velocity scaling method. The time step of MD was 0.9676 fs. The initial cluster structure for the MD simulation was a fcc-like cuboctahedron, unless otherwise stated. The atomic coordinates and Kohn-Sham energy were recorded every 100 steps up to 60000 steps and used as the learning data for machine learning. The atomization energy of the $Au_{13}^+$ cluster per atom ($E_A$) was defined as follows:

$E_A = \{13E_{KS}(Au) - E_{KS}(Au_{13}^+)\}/13,$

where $E_{KS}(X)$ refers to the Kohn-Sham energy of species $X$.

**Machine learning:** All machine-learning calculations in this study were based on the scikit-learn package, which is a collection of APIs for machine learning in Python.[28] To determine the relationship between the structure of a $Au_{13}^+$ cluster and its atomization energy, we considered two regression models: Gaussian kernel ridge regression (GKRR) and gradient boosting regression (GBR). We mainly used GKRR in this study. GKRR is a combination of the Gaussian kernel method and ridge regression with L2-norm regularization. Because of the flexibility afforded by the nonlinear character of the Gaussian kernel and ingenious kernel trick, GKRR efficiently finds relationships that ordinary linear regression fails to find.[12] GKRR contains two hyper-parameters σ and λ, which were optimized using a grid search, with exponential grids of 50 points between $10^{-5}$ and 1. We used $k$-fold cross-validation, where the training data set was divided into $k$-subsets, and the holdout method was repeated k times. In each case, one of the $k$-subsets was used as the test set and the other ($k$-1)-subsets were grouped together to form the training set. The mean score over $k$ trials was then calculated. We set $k$ = 10 and used the coefficient of determination, $R^2$, to score the model fitness.

GBR produces a regression model in the form of ensemble decision trees. It evolves the model in a stepwise manner by optimizing the loss function via its gradient. We used least square regression as the loss function (loss = "ls"). The learning rate at which the contribution of each decision tree shrank and the number of boosting stages to perform were set in a very conservative way to avoid overfitting (learning rate = 0.0001 and n_estimators = 25000). In addition, the maximum depth, minimum samples split, and maximum feature parameters of the scikit-learn gradient-boosting regression module (ensemble.GradientBoostingRegressor) were set to 5, 3, and 2, respectively.[28]

The data (features and atomization energies) were standardized. If the variance of a certain feature was significantly larger than that of the other features, there was a possibility of it dominating the objective function. It was also possible that the estimator might not correctly learn from the other features as expected. Although the standardization of data was not necessary for the GBR method because of its scale-invariance character, we also standardized its data to facilitate its comparison with the results of the GKRR method.

Results and Discussion

**Stable and metastable structures of the Au$_{13}^+$ cluster:** We prepared the test data to verify the present machine-learning approach. Gilb et al. showed 15 stable and metastable structures of Au$_{13}^+$ clusters obtained from quantum chemical calculations that were consistent with cross-section measurements, and reported the optimized atomic coordinates of the clusters.[24] Their coordinates were used as the initial geometries for our DFT calculations, and the recalculated optimized structures are shown in Figure 1. S0 was the most stable structure and M1–M14 were the metastable structures. A smaller number designating a cluster corresponds to a more stable structure in ref. 24. As stated, metal bonds have high flexibilities. Thus, it is possible to adopt diverse structures even in a 13-atom system, which is clearly observed in the figure. This makes the cluster suitable for verifying a new approach. Figure 2 compares the calculated relative energies of 14 clusters with the reported values.[24] Although the present calculations generally reproduced the results in ref. 24, the M3 and M9 clusters were less stable in this study than those in ref. 24. Calculations at the GGA level were carried out in ref. 24, although different GGA functionals were used. In addition, ref. 24 used Gaussian basis functions, unlike this study, which used planewaves. These differences might cause subtle differences in the results.

**Molecular-dynamics simulation:** To sample the learning data efficiently, it was preferable to set the simulation temperature such that the atoms could move actively. However, we found that when the temperature was too high the cluster was divided into two parts during simulation. Thus, we set the temperature to 1500 K, which was slightly higher than the melting point of Au (1337 K).[29] Starting the simulation from the fcc-like cuboctahedral initial structure, we recorded the time evolution of the cluster structure every 100 steps up to 60000 steps (58.05 ps). This MD simulation is shown in animation_W1, which is attached as a web-enhanced object (WEO [not included in arXiv version]). We found that the cluster structure greatly fluctuated during the first 20000 steps (19.35 ps), after which rotational motion was conspicuous. During the final stage of the simulation, it seemed that the cluster rotated with almost the same structure. Even if longer simulations were performed, it would not have been very useful for sampling the different structures. Thus, the time series data up to 60000 steps was used as the learning data. Although it was possible to artificially add forces that cancelled the rotational motion, it was unclear how the additional forces affected the prediction by machine learning, and so, we did not add such forces.

**Design of the descriptors for the Au$_{13}^+$ cluster**: Molecular structures were easily calculable descriptors. The interatomic distances were the first choice for such descriptors and the subject of the initial investigation. The distance also had the advantage of being invariant to rotational and translational motions. However, since there were $N!$ cases for indexing a molecule comprising $N$ atoms, the descriptors should be independent of atomic numbering. For such descriptors, we first considered those descriptors in which the interatomic distances were rearranged in the ascending order. Figure 3a shows the result of this case. The predicted values corresponding to a large atomization energy distributed relatively evenly above and below the 45°line, which represent perfect fitting. The predicted values corresponding to a small atomization energy seemed to be overestimated. The accuracy of the prediction was evaluated by the coefficient of determination ($R^2$). The $R^2$ of the training and test data were 0.741 and 0.255, respectively; neither was a good score.

To improve the results, it was necessary to examine descriptors that were different from interatomic distances. Although three- and four-body terms such as bond and dihedral angles were candidates for descriptors,[8,30] we stuck to using a simple expression based on the distance. We considered the distance between each atom and the mass center of the cluster. We called it the centroid distance. The centroid distances were also arranged in the ascending order, as in the case of interatomic distances. Figure 3b shows the result of the GKR model using only centroid distances as the descriptors. Obviously, the accuracy of the prediction for both the training and test data improved relative to the case where interatomic distances were used as the descriptors. The $R^2$ values of the training and test data were 0.864 and 0.579, respectively. We also show the case where both interatomic and centroid distances were used to extend the descriptors (Figure 3c). Although the fitness to the learning data improved ($R^2 = 0.941$), the fitness to the test data was somewhat reduced ($R^2 = 0.565$). We had to consider whether the extended descriptors had poor compatibility with the test data or they were overlearned because of the increased number of descriptors. In fact, as will be discussed below, the extended descriptors could improve the generalization capability by adjusting the number of features representing the interatomic distances.

**Feature importance:** The prediction of atomization energy was improved when centroid distances were used as the descriptors. To investigate this, we examined feature importance by constructing a GBR model. The least absolute shrinkage and selection operator (Lasso) is often used to examine the feature importance in a linear regression model. However, Lasso cannot be used for nonlinear regression models such as GKRR because it results in data selection instead of feature selection. In contrast, it was possible to calculate the importance of each feature in the GBR model. Figure 4 shows the results obtained by the GBR model using interatomic and centroid distances as the descriptors. The $R^2$ scores were 0.906 and 0.570 for the training and test data, respectively. These values were comparable to the scores from the GKRR model. However, the distribution of values predicted by the GBR model did not correlate much with the DFT values, unlike those of the GKRR model, in particular at large atomization energies. Thus, we mainly used the GKRR model in this study. Figure 5 shows the ranking of feature importance. A larger number indexing the feature refers to a longer distance. The symbol C was attached to designate the centroid distance. Most of the top-ranking features represented the centroid distances. The features C11, C12, and C13, which were particularly important, corresponded to a type of radius of the $Au_{13}^+$ cluster. In addition, C1, which was the third most important feature, represented the shortest centroid distance. Depending on the structure of the cluster, an atom could exist at the center of the cluster. In this case, C1 was almost zero. Therefore, C1 was a characteristic feature reflecting the shape of the cluster. Most of the features that had low importance levels represented short interatomic distances.

**Trimmed feature models:** The above results suggest that the features corresponding to short interatomic distances did not contribute much to the prediction of atomization energy. To confirm this, we examined how the $R^2$ score of test data changed with respect to the number of features by a sequential decrease by 5 from the feature representing the short interatomic distance. Near the peak $R^2$ score, the number of features was cut one by one. If all 78 features representing the interatomic distances were cut, the features comprised 13 corresponding to the centroid distances. The result is shown in Figure 6a. As the features corresponding to the short distance were cut, the $R^2$ score of the test data gradually increased, and the fitness to the test data was maximized when 40 features were cut. A scatter chart of the predicted values versus the DFT calculation values is shown in Figure 6b. Compared to Figure 3c, in which all the interatomic distances were considered, we found that the predicted values were significantly improved at points where the deviation was large from the DFT value. The $R^2$ scores for the training data with the 40-feature trimmed model and full-feature model were 0.893 and 0.941, respectively. In contrast, the $R^2$ scores for the test data with the trimmed model and full-feature model were 0.707 and 0.565, respectively. The trimming of the short-distance features significantly improved the fitness to the test data, but slightly worsened the fitness to the training data.

**Dependence on learning data:** Finally, we examined the dependence of prediction accuracy on the choice of learning data. We used 60000 steps of time-series data obtained from the MD simulation starting from the most stable structure (S0 in Figure 1) as another series of learning data. Except for the initial structure, the other conditions such as temperature were the same as those in the above MD simulation. This simulation is shown in animation_W2, which is also attached as a web-enhanced object (WEO [not included in arXiv version]). We observed that the cluster structure violently fluctuated during the first one-third and rotated with almost the same structure at the final stage. This behavior was in common with the first MD simulation. Note that the first learning data corresponded to the structure after 100 steps and the initial structure itself was not included in the learning data. We constructed a GKRR model including all 78 interatomic and 13 centroid distances as the descriptors. The result is shown in Figure 7. The $R^2$ scores for the learning and test data were 0.937 and 0.453, respectively. Although the score for the test data was somewhat worse because the prediction was not good for the M3 structure, the atomization energies of the remaining stable and metastable structures in Figure 1 were predicted well. Thus, we expect that this method did not depend much on the choice of learning data.

Summary

We found that the atomization energy of the $Au_{13}^+$ cluster could be predicted using interatomic and centroid distances as the descriptors. From the analysis of feature importance, we also found that features corresponding to the long centroid distance were important. Furthermore, by eliminating the features by about half from the shorter distance in the ascending order of interatomic distance, the fitness to the test data was markedly improved. These results indicate that the features corresponding to longer distances were important for describing the atomization energy of the cluster. This was contrary to the intuition that the chemical bond is local and the descriptors corresponding to shorter distances were expected to be important. The energy was estimated based on the cluster structure in this machine-learning approach. It could thus be interpreted that the emphasis was placed on the descriptors corresponding to longer distances, which were more likely to reflect the shape of the entire cluster. In this study, we focused on a system containing one element, but we hope that this method would be effective even when it contains many elements. For example, in binary compounds consisting of two elements A and B, this method might be applicable by separately considering the distance of each pair AA, AB, and BB.

**Supporting Information**. The following files, provided as animation_W1 and animation_W2, are available free-of-charge as a web-enhanced object (mpeg) corresponding to MD simulation starting from cuboctahedron (W1) and stable structure S0 (W2).

**Notes**

The authors declare no competing financial interests.

Figure captions

Figure 1: Ball-and-stick models of stable (S0) and fourteen metastable (M1-M14) structures of the $Au_{13}^{+}$ cluster obtained from DFT calculations.

Figure 2: A comparison of the present calculation (red diamonds) with the calculation in ref. 24 (blue circles) with respect to the relative energies of $Au_{13}^{+}$ clusters. Zero energy corresponds to the energy of the S0 structure.

Figure 3: Machine learning and prediction of $Au_{13}^{+}$ atomization energy by GKRR models with standardized data: (a) interatomic distances, (b) centroid distances, (c) interatomic and centroid distances used as the descriptors. Optimized hyper-parameters in GKRR model: (a) $\alpha = 0.30888$ and $\gamma = 0.009103$, (b) $\alpha = 0.19307$ and $\gamma = 0.047149$, and (c) $\alpha = 0.07543$ and $\gamma = 0.007197$. The blue and red circles correspond to training data and test data, respectively. Note that the axes are standardized.

Figure 4: The same as Figure 3c, but the prediction was done using the GBR model. All 78 interatomic and 13 centroid distances were included as the descriptors.

Figure 5: Ranking of feature importance of the GBR model shown in Figure 4. The indexing was assigned in the ascending order of distance and the symbol C designates centroid distance.

Figure 6: (a) $R^2$ scores for the training (blue circles) /test (red diamonds) data versus the number of reduced features representing interatomic distances in the ascending order of distance. (b) The same as Figure 3c, but the prediction was done using the GKRR model after cutting the 40 shortest interatomic distances.

Figure 7: The same as Figure 3c but obtained from different MD simulations starting from the most stable structure S0 in Figure 1. All interatomic and centroid distances were included as the descriptors.

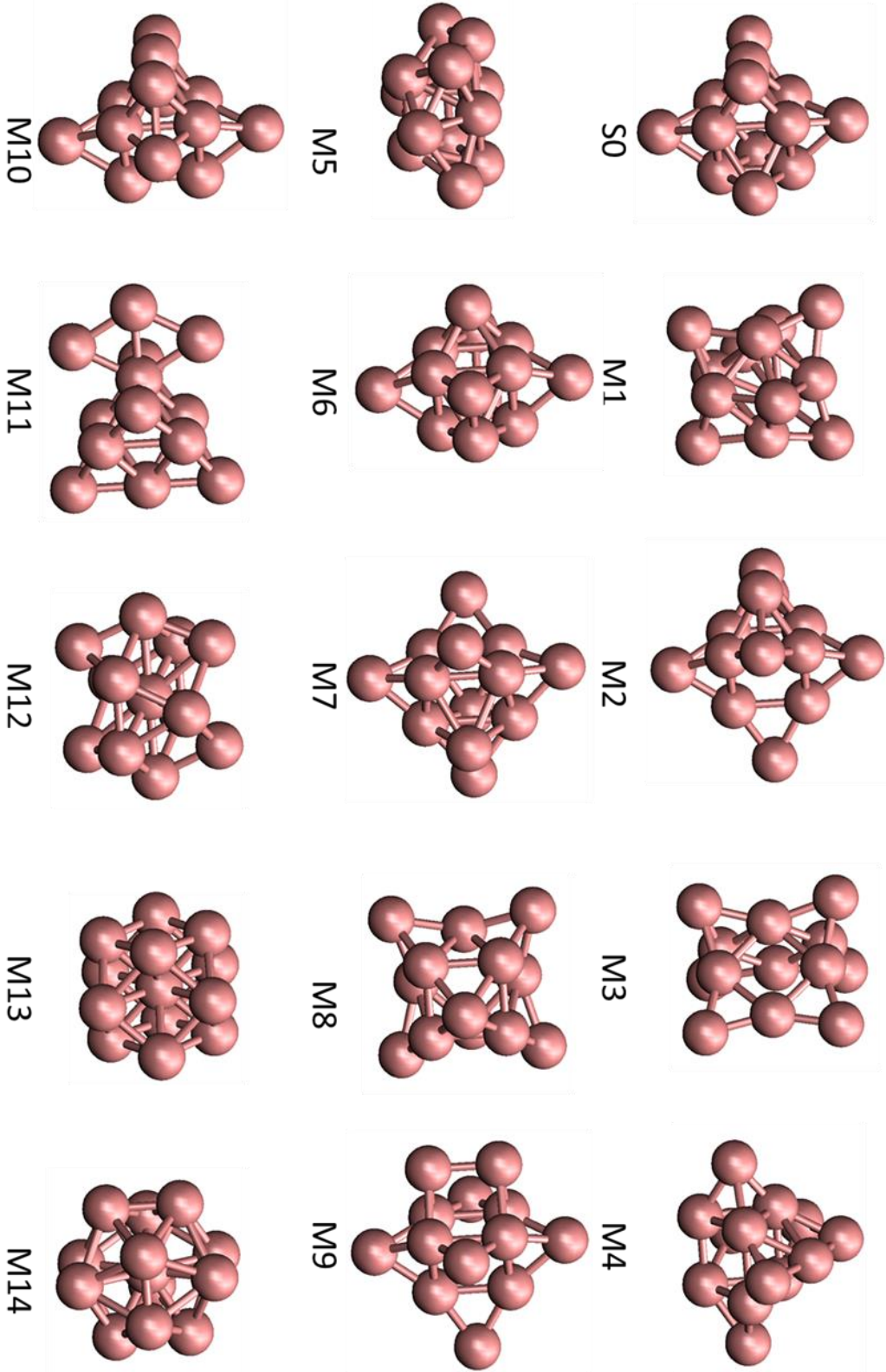

Figure 1

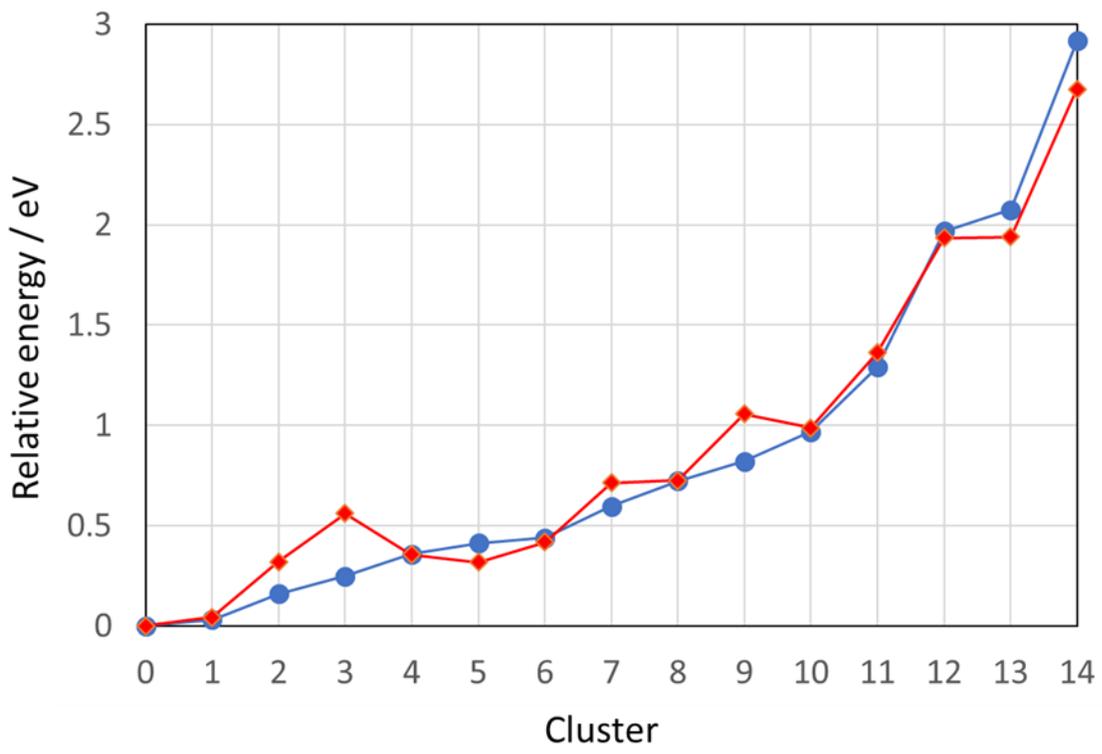

Figure 2

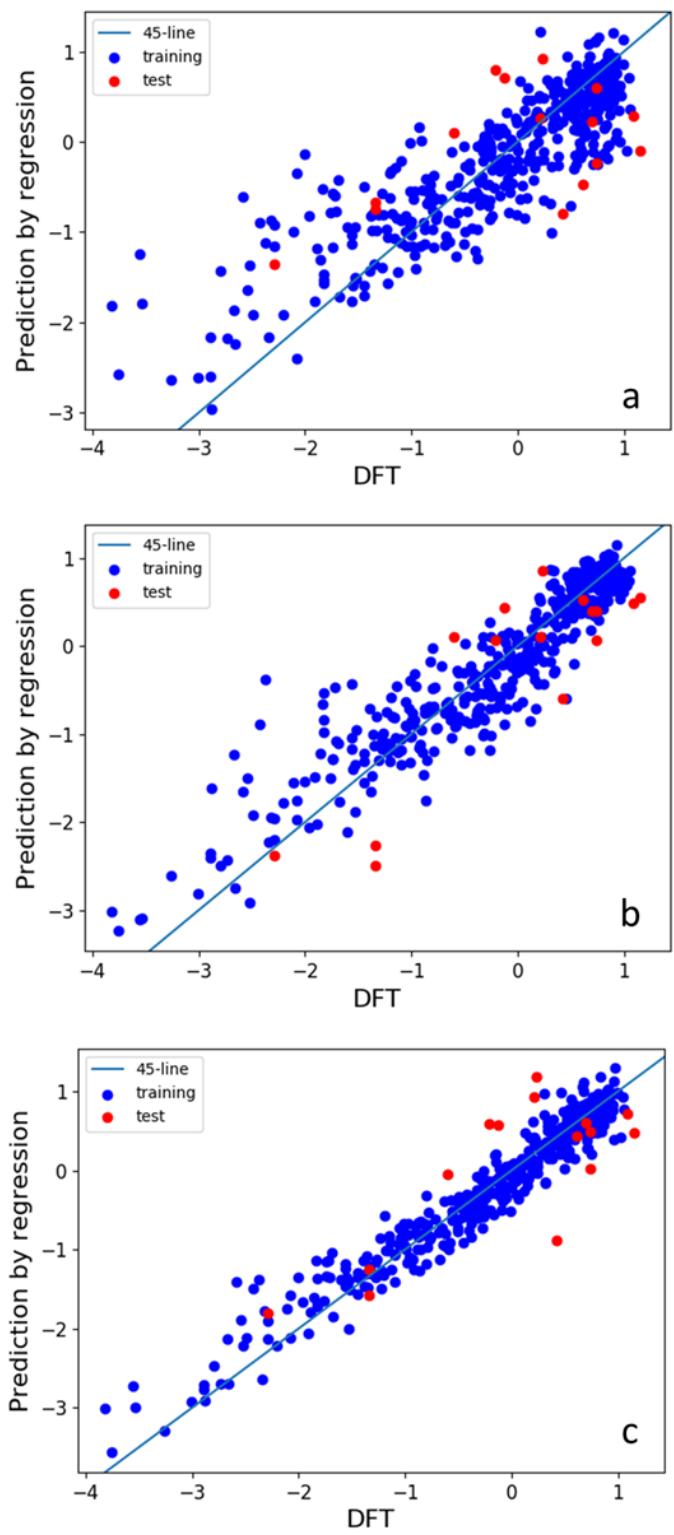

Figure 3

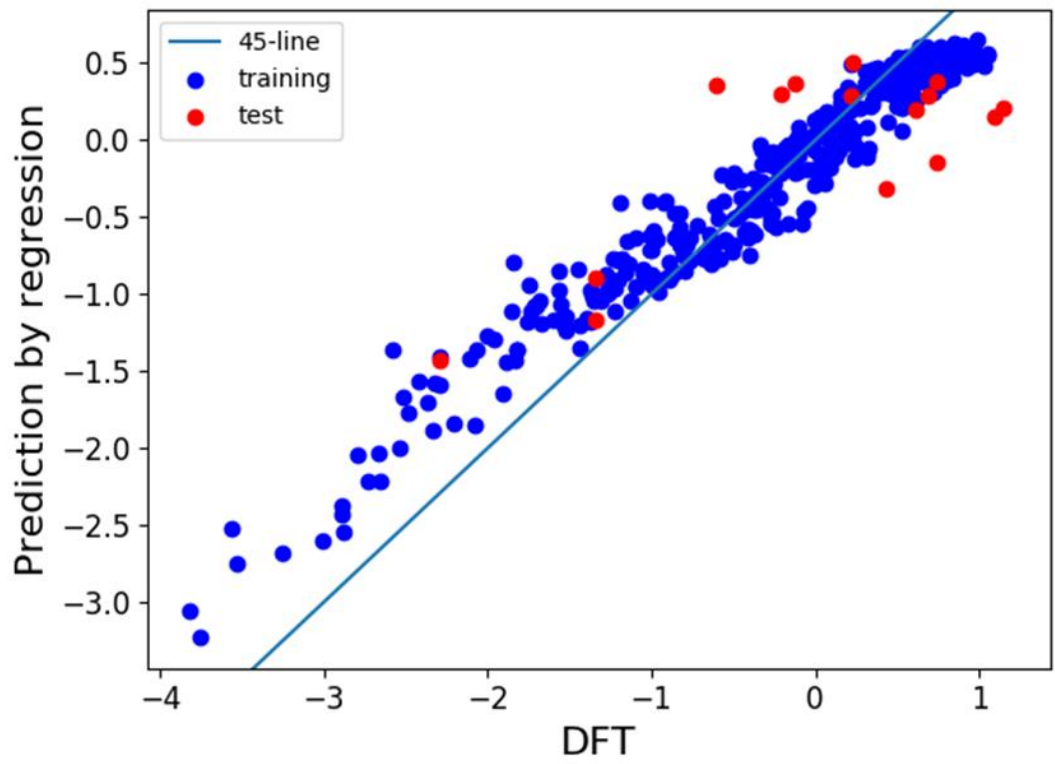

Figure 4

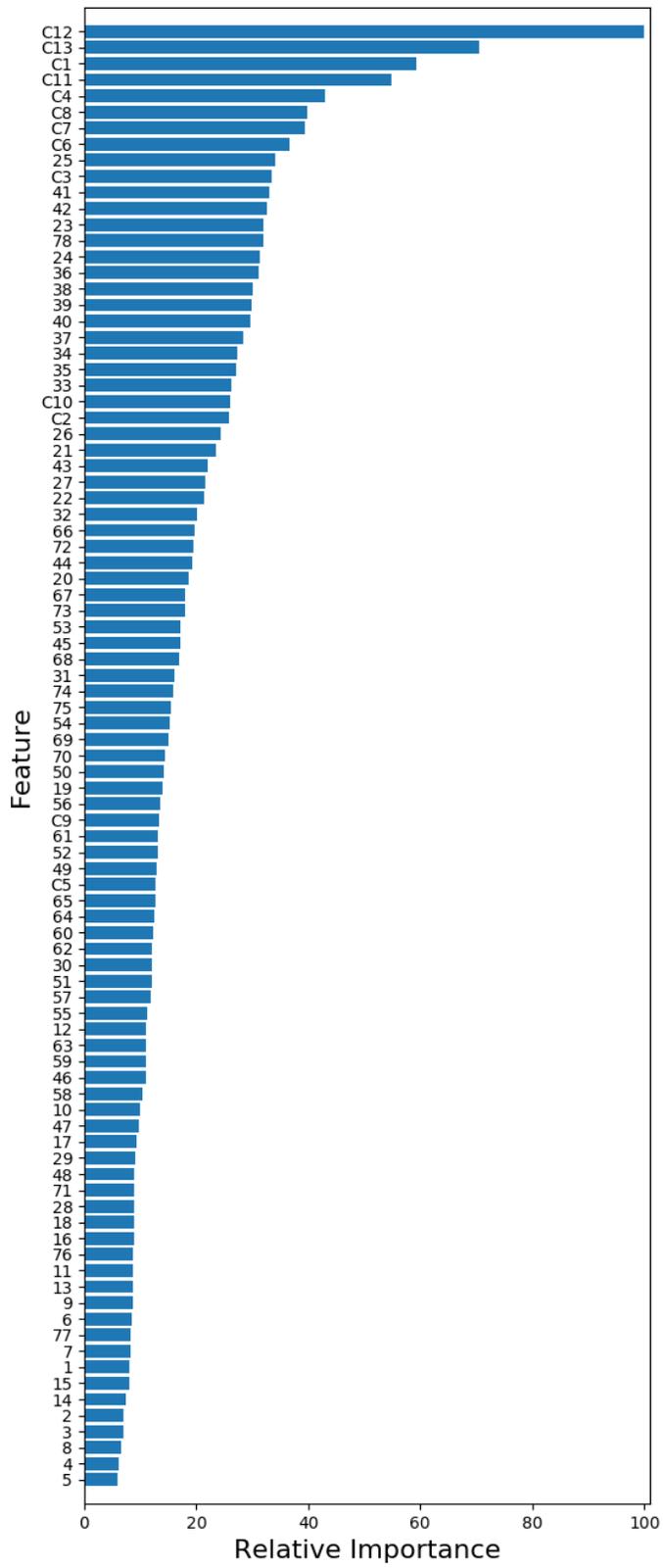

Figure 5

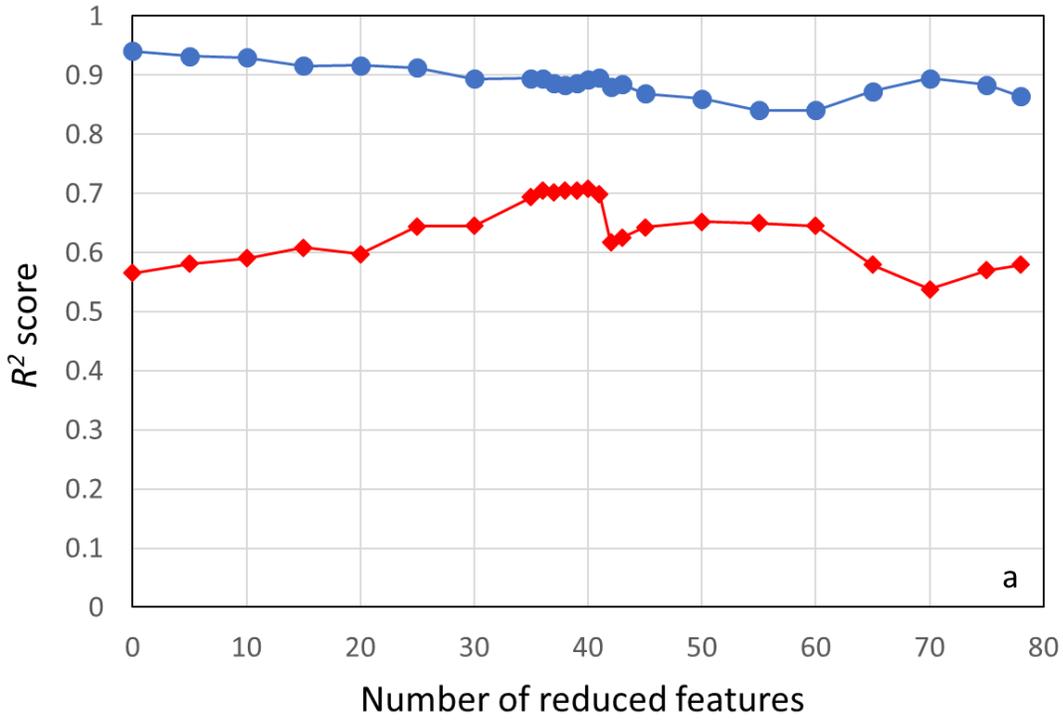

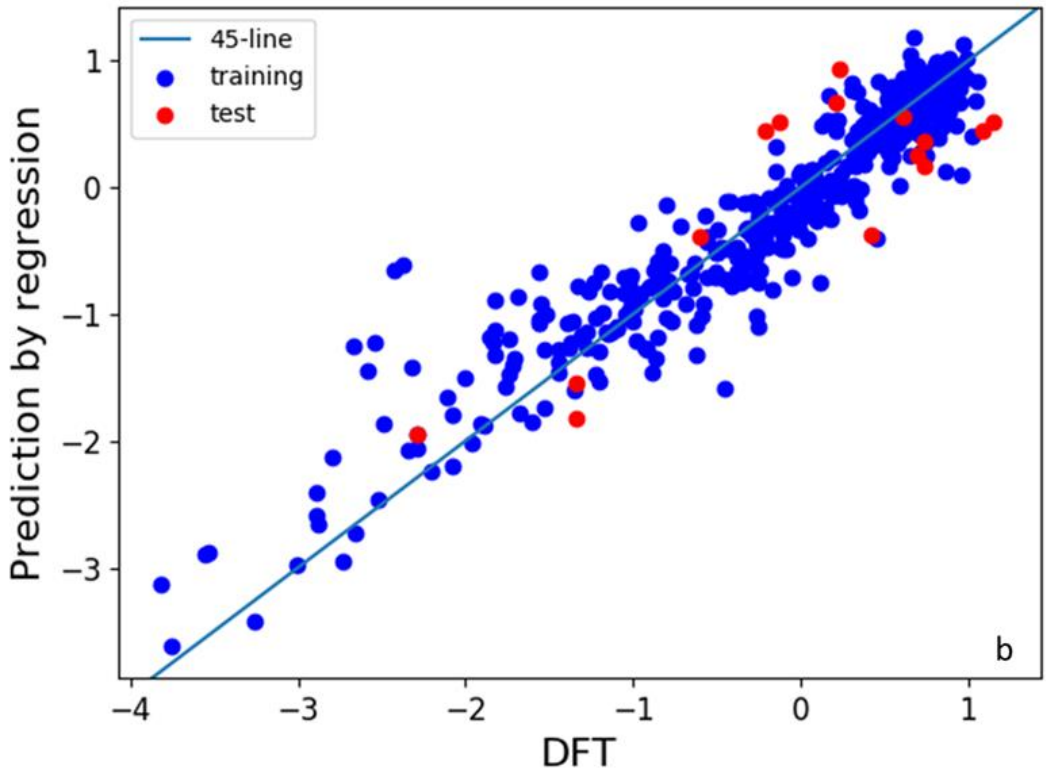

Figure 6

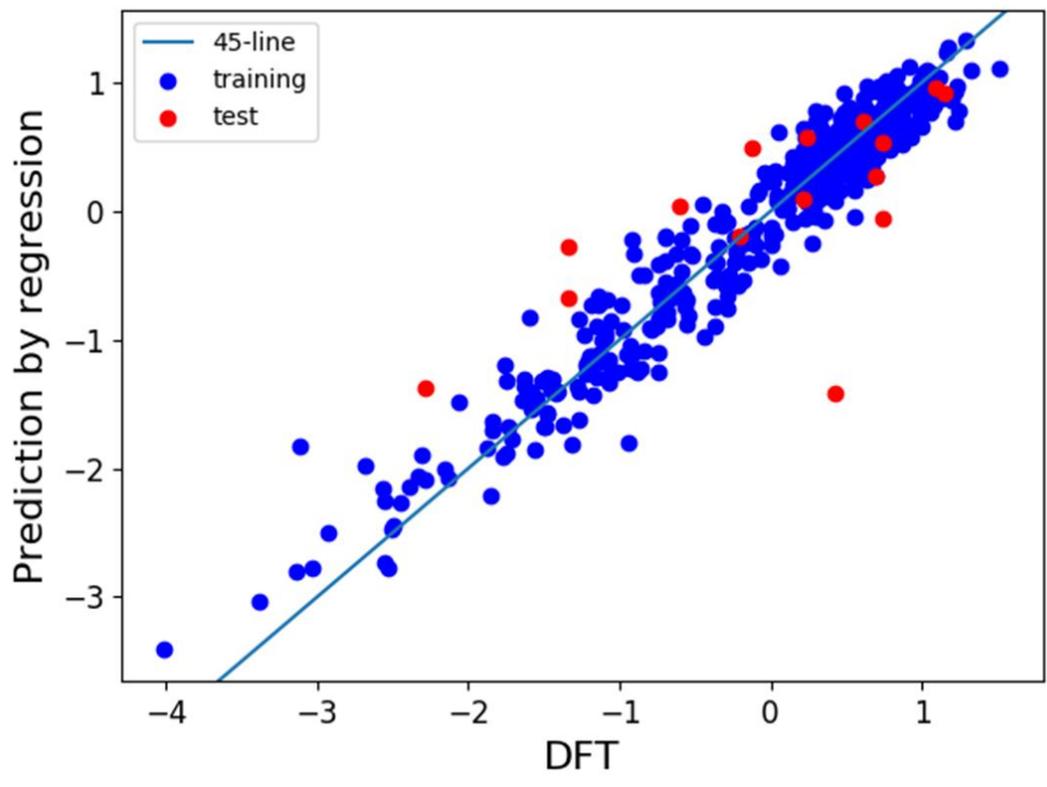

Figure 7